\documentclass[fleqn,10pt]{olplainarticle}
\usepackage{tikz-cd}
\usepackage{hyperref}
\usepackage{graphicx,color}
\usepackage{amssymb,amsmath, bm}
\usepackage{mathrsfs}
\usepackage{ragged2e}
\usepackage{tkz-graph}
\usepackage{adjustbox}
\usepackage{tikz}
\usepackage{float}
\usepackage{caption}
\usepackage[all]{xy}
\usetikzlibrary{arrows,decorations.pathmorphing,automata,backgrounds}
\usetikzlibrary{backgrounds,positioning}
\usepackage{transparent}
\usepackage{graphicx,import}
\usepackage{subcaption}
\usepackage[utf8]{inputenc}
\DeclareUnicodeCharacter{0301}{\'{}}  
\usepackage{setspace}
\usepackage{ulem}

\newtheorem{theorem}{Theorem}[section]
\newtheorem{example}[theorem]{Example}

\numberwithin{equation}{section}

\title{From Shape to Species: The Use of Persistent Homology for Identifying Quantitative, Taxonomically Informative Morphological Traits in \textit{Acropora} Corals}

\author[1]{Lilith Eringfeld}
\author[1]{Sophie Raynor}
\author[2,3]{Yossi Bokor Bleile}
\author[1,4]{Tom C. L. Bridge}
\affil[1]{College of Science and Engineering, James Cook University, Townsville, Australia}
\affil[2]{Institute of Science and Technology Austria, Klosterneuburg, Austria}
\affil[3]{School of Mathematics and Statistics, The University of Sydney, Sydney, Australia}
\affil[4]{Biodiversity and Geosciences Program, Queensland Museum Tropics, Townsville, Australia}

\keywords{Persistent Homology, \textit{Acropora}, Coral Morphology, Coral Taxonomy}

\begin{abstract}
Molecular phylogenetics is fundamentally altering our understanding of the taxonomy and systematics of scleractinian corals. Traditionally, coral taxonomy has relied on the comparison of morphological traits of the coral skeleton, but molecular data has shown that in many cases diagnostic traits are either convergent or so morphologically variable as to be uninformative. Although phylogenetically informative characters have been identified in some clades, we still lack quantitative, repeatable methods for identifying taxonomically informative morphological traits for most coral species either in the field or in museum collections. Geometric and Topological Data Analysis (GTDA) combines concepts from geometry and topology to understand complex, high-dimensional, and noisy datasets, and can detect features that familiar methods overlook. This novel approach makes GTDA and persistent homology a potential method for quantifying coral morphology in a taxonomic context. Here, we investigate the utility of GTDA to identify taxonomically informative traits in a complex of tabular \textit{Acropora} coral species delineated a priority using molecular phylogeny. The results show that taxonomically informative topological information can be extracted from in-situ coral images and that there is great potential in further refining this method. This study is a first exploration into the potential of persistent homology in supporting the development of quantitative, repeatable methods for delineating \textit{Acropora} species from digital images. Given the key role \textit{Acropora} plays on tropical reefs across the Indo-Pacific and their importance for reef restoration and conservation, the development of repeatable methods to effectively identify species within the complex genus \textit{Acropora} would be of significant benefit to research and management of coral reef ecosystems.

\end{abstract}

\begin{document}

\maketitle
\thispagestyle{empty}
\begin{figure}[H]
\centering
\includegraphics[width=0.8\textwidth]{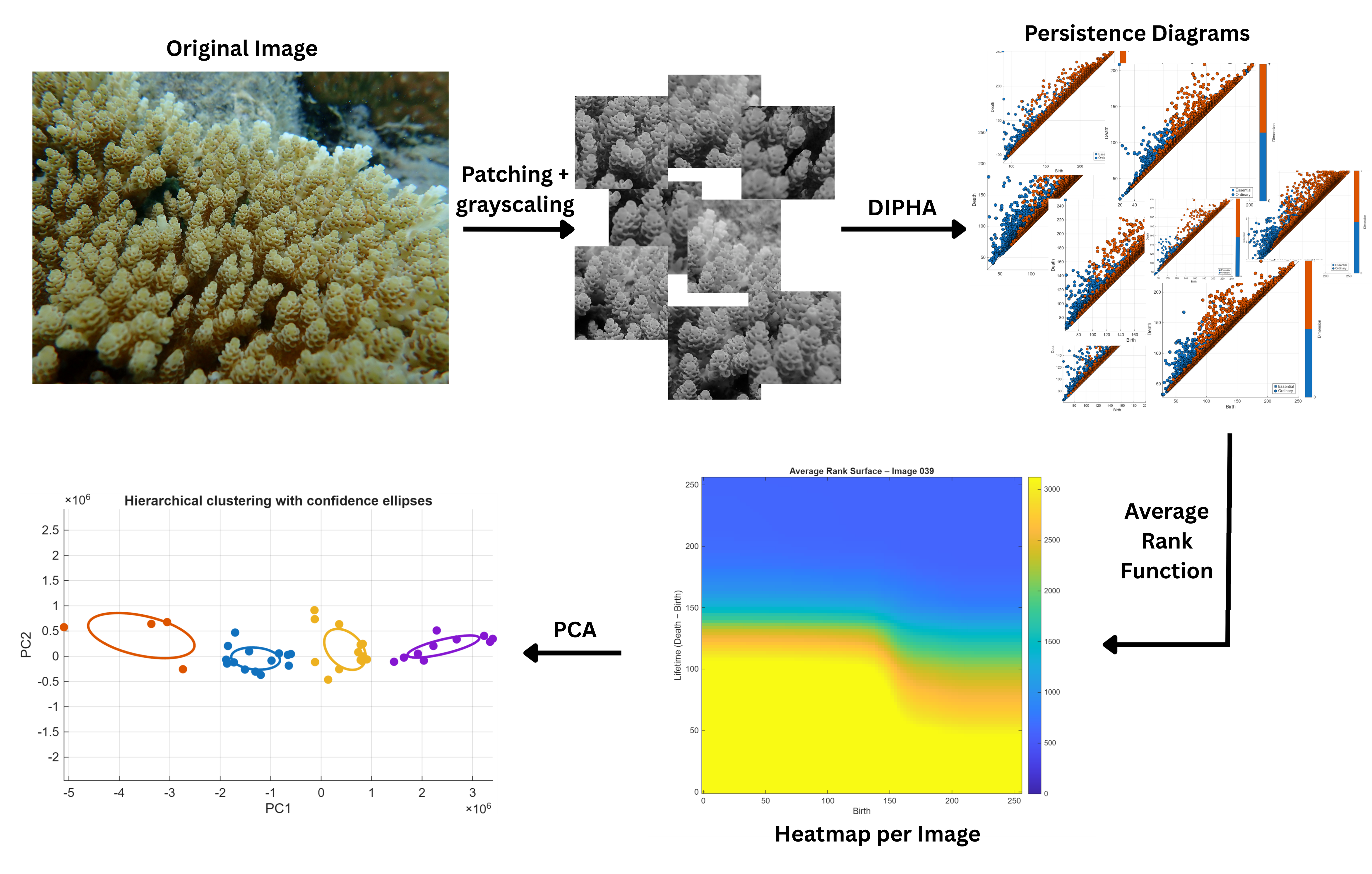}
\caption{Schematic showing the method pipeline to process the images into statistical values.}
\label{fig:method}
\end{figure}

\section{Introduction}

Robust taxonomy that accurately reflects the diversity and evolutionary history of a clade underpins virtually all aspects of biological, ecological, and conservation science and management \citep{thomson2018taxonomy,sandall2023globally}. For over two centuries since the formalization of the binomial system of classification by \cite{Linnaeus1758}, species identities and systematic relationships were inferred almost exclusively through comparative analysis of morphological traits. However, the advent of molecular phylogenetics in the 1970s provided a new tool to examine evolutionary relationships and has fundamentally altered our understanding of the systematics and evolution of many groups across the Tree of Life. Specifically, molecular phylogenies regularly illuminate the fact that morphological characters traditionally used to delineate species or higher taxonomic groups are homoplasies and not indicative of close evolutionary affinities \citep{Moore1997convergence,Mahler2013convergence}. Consequently, robust delineation of lineages that reflects evolutionary history requires the identification of morphological characters that are phylogenetically informative; that is, that they are diagnostic of a particular clade.

Hard corals are the dominant ecosystem engineers of tropical reefs, the most diverse marine ecosystems on Earth. However, the abundance of corals on reefs is declining rapidly due to climate change and other anthropogenic activities \citep{Hughes2017global}. Despite their ecological importance and well-documented declines in abundance, we still lack the capacity to accurately identify coral species. Consequently, little information is available on key questions such as species abundances, geographic ranges, population trajectories, or extinction risk. This knowledge gap is driven primarily by the fact that some coral species exhibit extensive plasticity at the macromorphological level  \citep{Todd2008plasticity}, while others can appear very similar despite deeply divergent evolutionary histories \citep{Fukami2004taxonomy}; in other words, in some cases a single species can vary extensively in morphology, while in others two distantly related species can appear very similar.

For most of their taxonomic history, corals have been classified based on morphological traits of their aragonite skeletons. In most cases, the authors of the taxonomic monographs of the 19th century that described coral species and placed them into their respective higher-level taxonomic groupings were based in museums in Europe or North America, and few had ever seen corals in the field. Even those that did participate in scientific expeditions were still limited by the technology of the day, and corals could only be collected either from very shallow reef flats at low tide or by destructive means such as dredging. The advent of SCUBA diving in the 1970s as the primary method for studying corals enabled researchers to access a much wider range of coral reef habitats and observe corals in situ, and popularized the idea that much of the variation observed between species was actually intraspecific variation within a single species depending on the environment \citep{veron1976scleractinia}. The influential series of monographs ‘Scleractinia of Eastern Australia’ \citep{veron1976scleractinia, veron1980scleractinia, veron1982scleractinia, veron1984scleractinia, veron1977scleractinia} were the first taxonomic revisions based on extensive collections of corals collected on SCUBA. The idea that morphological variation between species was due to environmental plasticity, combined with the idea that most coral species had wide geographic ranges spanning much of the Indo-Pacific \citep{stehli1971diversity,veron2000}, led to the synonymy of many nominal species on the basis that they represented intraspecific variation within morphologically variable and geographically widespread species. 

The taxonomic scheme developed in the Scleractinia of Eastern Australia series has underpinned coral science for much of the last half a century. However, beginning in the 1990s, molecular phylogenies of reef corals began challenging the traditional morphological taxonomy and revealed that these classifications were incongruent with systematic relationships inferred using morphology at all taxonomic levels \citep{romano1996evolution,Fukami2004taxonomy,fukami2008mitochondrial}. While many of the macromorphological characters traditionally used to define species and genera, such as corallite integration and mode of calicular budding, were shown to be homoplasies, integrated approaches combining detailed morphological analysis of micromorphological and microstructural elements of the coral skeleton were congruent with the systematic relationships identified by molecular phylogenies \citep{budd2009searching, budd2010rethinking, budd2011corallite, budd2012taxonomic, huang2014towards, huang2014taxonomic, huang2016taxonomic}. These results indicate that many clades do indeed possess phylogenetically informative traits, but identifying them requires examining which traits are conserved within clades delineated by molecular phylogenies.

The studies outlined above underpinned a comprehensive revision of the taxonomy of reef corals at all taxonomic levels. However, phylogenetic analyses capable of resolving species-level relationships in the most taxonomically diverse coral taxa, such as Acropora, have only recently been developed \citep{cowman2020enhanced, ramirez2022quantitative, bridge2024tenuis, rassmussen2025tables}. Molecular-based species delimitation techniques, combined with qualitative comparisons of the morphology of coral skeletons from sequenced specimens, museum collections, and type material, have enabled taxonomic revisions with Acropora and revealed that both taxonomic diversity and endemism have been substantially underestimated. Despite significant advances in the capacity for researchers with taxonomic expertise to use the results of this work to identify specimens in the field, we still lack the means to quantitatively describe the skeletal morphology of a coral colony. Accurate identification of most Acropora colonies to species level requires genetic sequencing, which is costly and takes considerable time, and is therefore not practical in many situations. Consequently, our capacity to identify species in the field, in museum collections and, importantly, to communicate how to identify species to others, remains limited. This issue is compounded by the fact that coral surveys are increasingly conducted using automated image classifications (such as ReefCloud), and poor training of a model will result in poor results. Consequently, there is an urgent need to quantitatively describe the morphology of coral skeletons in a way that allows species to be delineated in the field and in collections.

\begin{figure}[ht]
\centering
\includegraphics[width=1\linewidth]{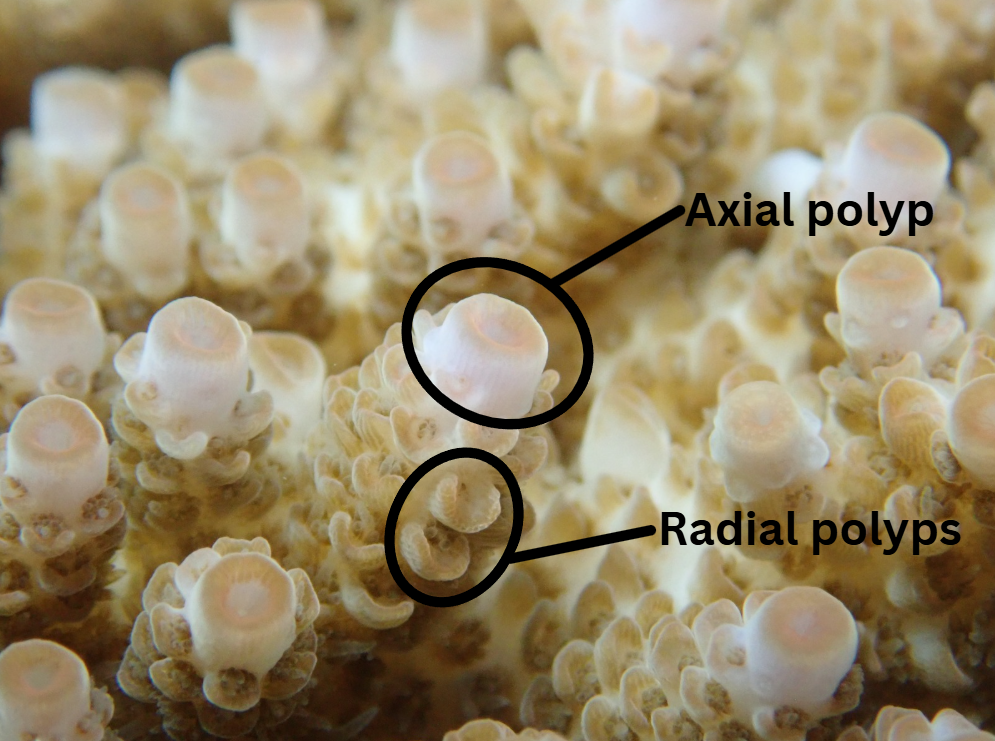}
\caption{Close-up of \textit{A. hyacinthus} showing the distinct, bigger axial polyp at the tip of the branch and the smaller radial polyps around the thick of the branch. From \textit{QMT CoralBank}.}
\label{fig:polyps}
\end{figure}
\newpage
This gap is especially consequential for \textit{Acropora}, the most species-rich and abundant coral genus worldwide \citep{hoeksema2026worms} and among the most vulnerable to Anthropocene disturbances \citep{darling2012lifehistory}. As the primary reef-building genus, its diversity remains understudied relative to its ecological role \citep{plaisance2011diversity}, and underestimating that diversity carries real conservation costs \citep{grupstra2024cryptic}: common species are easily overestimated while endemic species may be overlooked \citep{devantier2017richness, dornelas2008multiple, bridge2020incongruence, bridge2024tenuis}. This affects our ability to assess the impact of biodiversity losses \citep{richards2013regional, balmford2005trends, possingham2002limits, huang2024loss, richards2013molecular} and ultimately weakens reef management \citep{bortolus2008error}.

\begin{figure}[ht]
\centering
\includegraphics[width=1\linewidth]{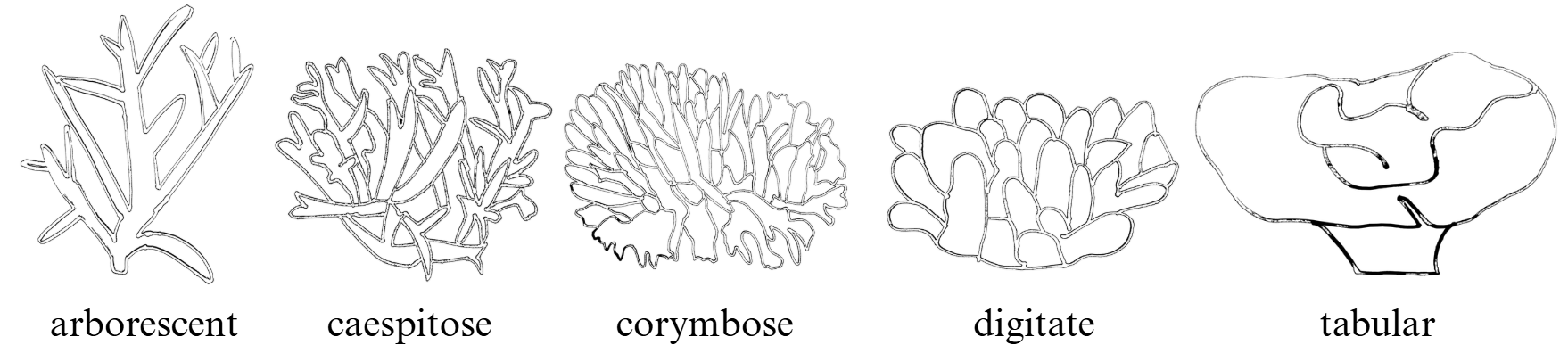}
\caption{Schematic showing the different growth forms of \textit{Acropora}.}
\label{fig:morphologies}
\end{figure}

\textit{Acropora} is often distinguished by features that are easy to see but hard to quantify, such as the size and shape of the axial and radial corallites (Fig. \ref{fig:polyps}), and the gross morphology of the colony, which can range from tabular to corymbose (Fig. \ref{fig:morphologies}). Earlier coral morphology studies tried to capture this variation with branching indices \citep{kruszynski2007method, alamsyah2024pocillopora}. However, these indices only partly describe morphological differences, since no exact criteria define when a branch splits or when two branches merge. As a result, these morphological signals have been underused, despite holding valuable structural information.

To address this gap, we propose the use of persistent homology, a tool from geometric and topological data analysis. Persistent homology captures morphological features - such as (topological) connectivity and 'loopiness' - that are not described by traditional metrics like length and angle. When applied to 2D grayscale images of \textit{Acropora} colonies, persistent homology offers a way to quantify morphological features directly from photographs, without tissue sampling or specialised skeletal preparation. The persistence features of each image may be summarised using a rank function, and we can apply principal component analysis (PCA) to find a representation that captures how the rank functions relate to one another, and test whether \textit{Acropora} species separate into distinct groups in this topological feature space. We hypothesise that, when applied to suitable images, persistent homology can recover species-level morphological differences from image data alone, providing a scalable and non-invasive method to quantify morphological traits of the coral skeleton that can be used to identify species. Moreover, persistent homology applied to existing image collections provides an efficient and low-cost method for identifying key principles for future data collection and pre-processing.

\section{Methods and Materials}

Our pipeline is as follows (see Fig.~\ref{fig:method} for a visual representation): original images are patched randomly, resulting in 30 equal-sized patches. These patches are then converted to grayscale and inverted, so that the foreground and tips of the coral branches appear dark. From the persistent homology (see Section~\ref{ssec ph} below), we compute the rank function. This provides a summary of the persistence, visualised as a heatmap, and enables us to compute the mean rank function for each image. Finally, PCA is performed on the mean rank functions. 

\subsection{Persistent Homology}\label{ssec ph}

Geometric and topological data analysis (GTDA) is a relatively new field that addresses questions about the shape and structure of a data set \citep{munch2017users}. GTDA is powerful at detecting subtle patterns that conventional approaches overlook \citep{lemus2025topo}. Based on principles from algebraic topology (a branch of mathematics), GTDA concisely summarises large and complex data sets, offering a new perspective on the shape of data \citep{amezquita2020shape}. Persistent homology is a tool from GTDA that effectively captures local structures while retaining global context \citep{katar2025skin}. 

\begin{figure}[H]
\centering
\includegraphics[width=0.9\linewidth]{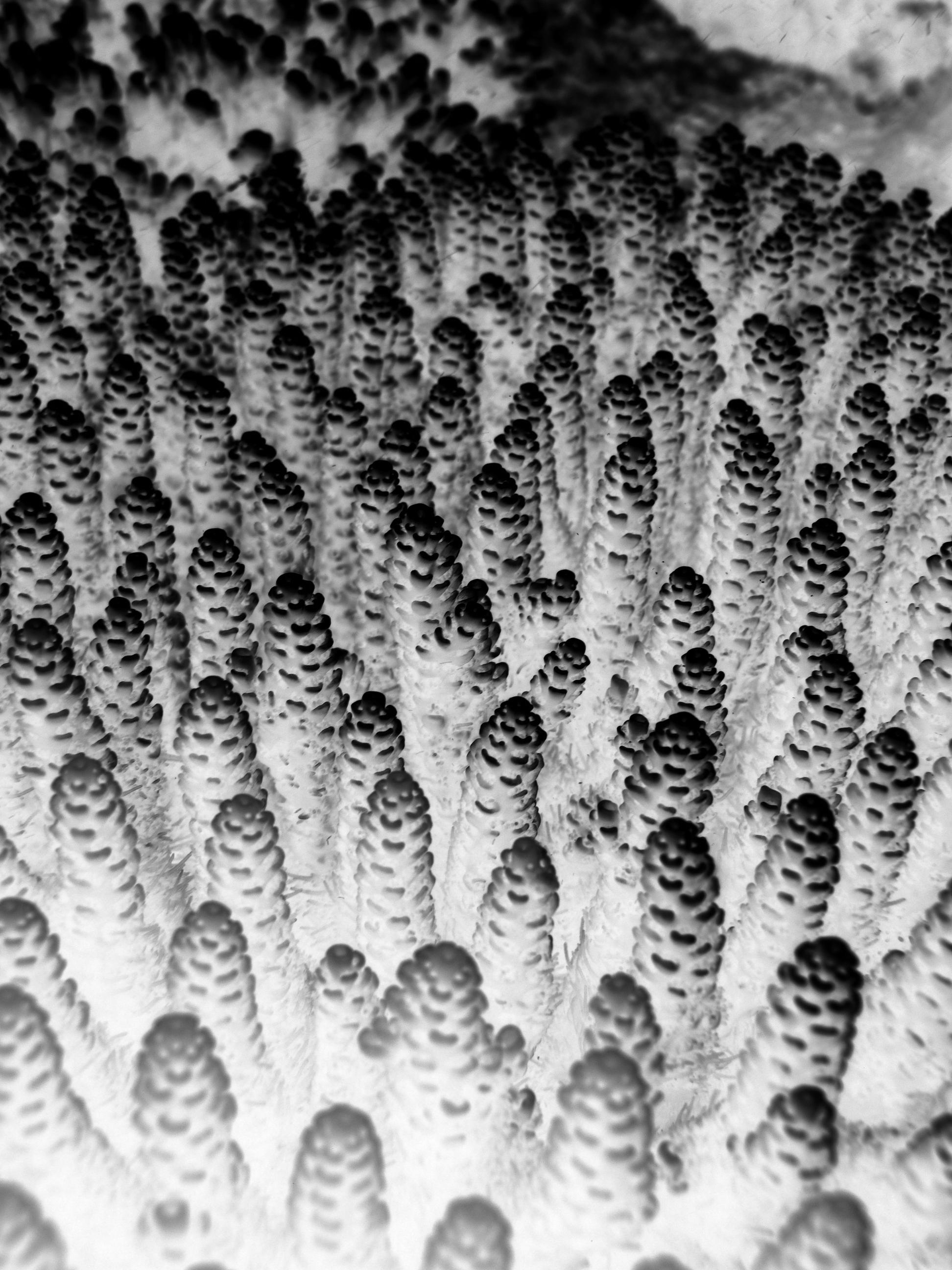}
\caption{Inverted grayscale image of \textit{Acropora sp.}. The tips of the branches and corallites/polyps are a darker tone compared to the base of the branches. From \textit{QMT CoralBank}.}
\label{fig:tone}
\end{figure}

Before outlining persistent homology, as applied to coral images in this work, we discuss the features visible in an inverted grayscale image of \textit{Acropora}. Looking at Fig.~\ref{fig:tone}, which shows \textit{Acropora sp.} in situ, the tips of the coral branches are generally much darker than the denser coral base. In particular, the axial polyp clearly sticks out, almost appearing black. Even the radial polyps around the coral branch appear in dark crescents, becoming lighter towards the base of the branch. The outline of the branches to the left and right tends to be darker as well, marking the vertical expansion. These changes can be useful for understanding the \emph{morphology} of the coral; however, they are not well defined by traditional morphological metrics. To make a summary of such morphological features, we turn to concepts from GTDA, in particular \emph{persistent homology}.

Persistent homology summarises the morphology of an object by tracking changes in (homological) features as a given parameter is varied \citep{carlsson2009topology}. In our setting of 2D grayscale images, this parameter is pixel intensity. Persistent homology tracks two types of features: connected components (represented by classes in degree $0$ homology, denoted $H_0$) and loops (represented by classes in degree $1$ homology, denoted $H_1$). These homology classes can be used as a measure of connectivity and arrangement of features in the image. 
Precise mathematical definitions for these features are available; we refer interested readers to the work of \cite{edelsbrunner2010computational}.
To keep the focus on the biological applications, the formal algebraic construction is beyond the scope of this study.

Our method is applied to 2D grayscale images composed of pixels. Each pixel is assigned a grayscale value $I_{ij} \in [0,255]$, where 0 = black and 255 = white. Persistent homology works by tracking the homology classes through a process called filtration. In our method, the filtration process first considers a darker tone as we move towards white, so that pixels are progressively evaluated by their intensity (Fig.~\ref{fig:filter}). For each grayscale value $t$, we may consider the black and white image whose pixels with grayscale value greater than $t$ are white, and those with grayscale value less than or equal to $t$ are black. By continuously increasing the threshold value $t$, we obtain a sequence of black and white images, filtered by tone, whose evolving features are recorded by persistent homology. 

The pixel value at which a feature -- a connected component or loop -- appears 
is the \textit{birth time} $b$ of the corresponding homology class. Similarly, the pixel value at which it 
disappears (either through merging of components or filling of loops) is the \textit{death time} $d$. The difference between birth and death $d-b$, or the lifetime of a homology class, is referred to as its \textit{persistence} \citep{edelsbrunner2010computational}. Homology classes are then classified based on their birth and death times, creating a summary of topological features for the entire image.


This summary is recorded as a \textit{persistence diagram} (Fig.~\ref{fig:diag}), a collection of birth and death points in $\mathbb{R}^2$ \citep{moon2023persistent}. The persistence diagrams visualise the persistence of a homology class by featuring death on the y-axis and birth on the x-axis. The further away a data point is from the diagonal $y=x$, the longer its persistence is \citep{cohen2005stability}. Persistence diagrams can be a first indicator of certain features of the image, such as branch density or the prevalence of small, short-lived loops corresponding to fine surface texture.

\begin{example}
A step-by-step filtration helps visualise how these two homology classes may appear (Fig.~\ref{fig:filter}). At $t=14$ begins with the birth of 5 connected components (the black pixels) and one big loop. When $t=20$, two more connected components are born. At $t=29$, the two connected components on the left edge merge into one; simultaneously, a new connected component is born to the right of the structure. As the new connected component only touches diagonally with the next connected component, they remain separate. At $t=57$, all existing connected components merge into one, and a new one is born towards the centre of the grid. As the connected component grows, its shape closes around three separate regions of background pixels, sealing each one off from the rest of the image. Each enclosed region constitutes a hole, and the boundary path encircling each hole corresponds to a newly born loop (i.e., a class in $H_1$). When $t=110$, two new connected components are born in the lower right corner. One of them merges into the big connected component, which therefore gives birth to a new loop when $t=144$. At $t=156$, the loop in the lower left corner shrinks, but no change in the number of features is evident. The group of pixels must be fully closed off by the connected component to give rise to a loop. A single diagonal (8-)connection can sufficiently seal off the pixels without direct edge-adjacency. Therefore, the boundary path around the white pixels cannot be contracted to a single point; hence, the two white pixels in the upper-left corner belong to the same loop. This may not be intuitive; hence, it is a good reminder that GTDA analyses the shape of the data, not the shape we see drawn. When $t=220$, the connected components grow and merge, killing two loops.

\begin{figure}[htbp]
\centering
\includegraphics[width=0.5\linewidth]{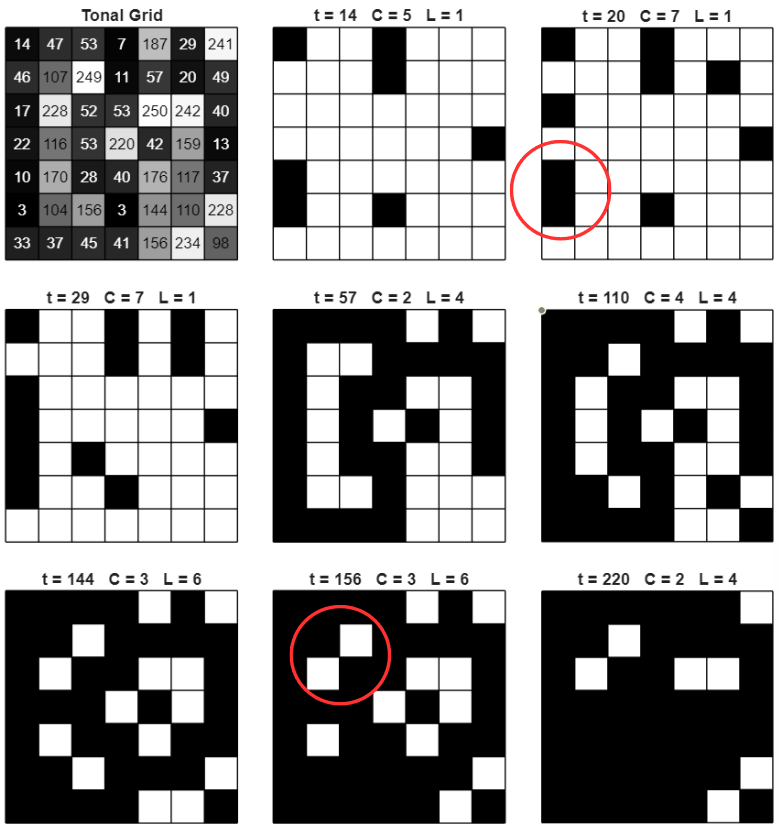}
\caption{A filtration considered in ascending order for each threshold value $t \in 0,\dots, 255$ from a tonal grid visualising how loops and connected components appear. t = threshold value, C = number of connected components, L = number of loops. The red circle when $t=20$ identifies a connected component, and when $t=156$, a loop.}
\label{fig:filter}
\end{figure}

\begin{figure}[htbp]
\centering

    \centering
    \includegraphics[width=\linewidth]{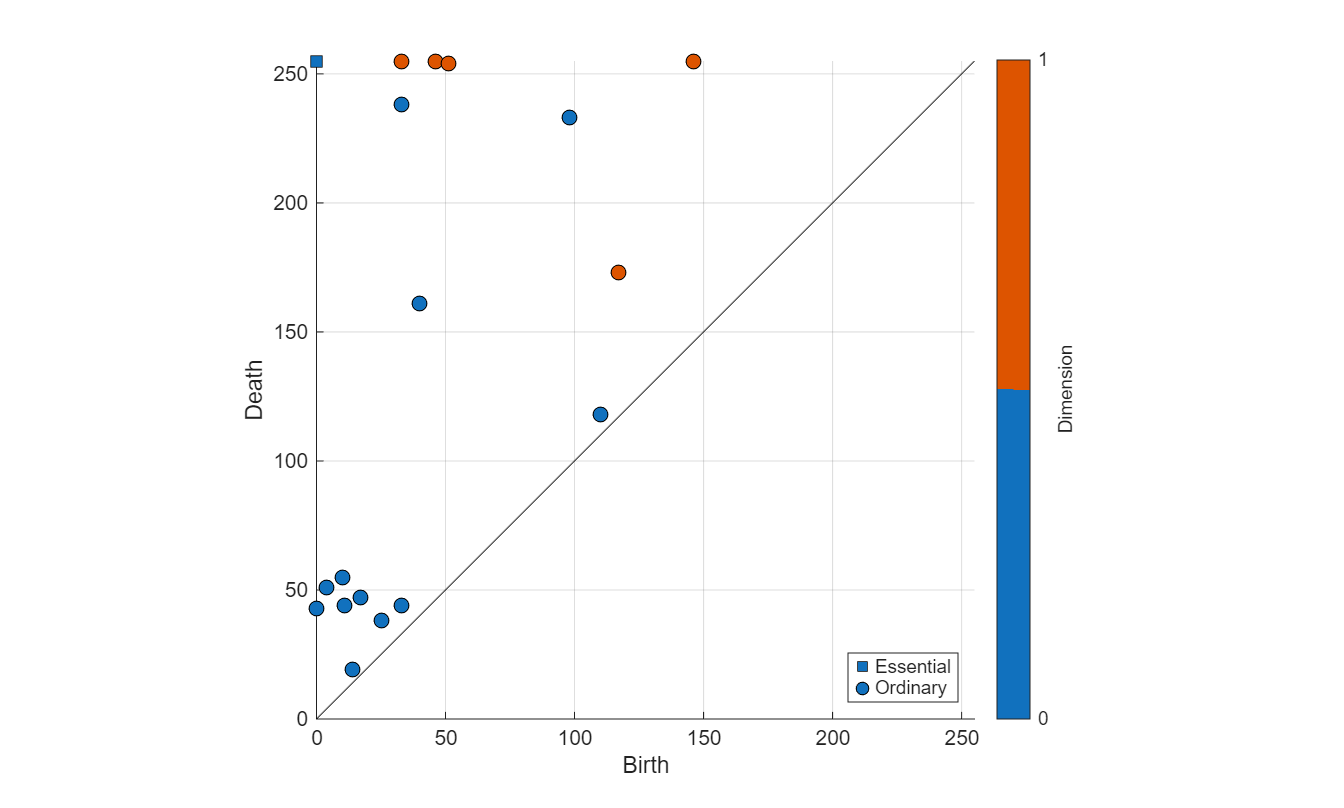}
    \caption{Persistence diagram.}
    \label{fig:pd}

\caption{Illustration of the persistence diagram resulting from the filtration in Fig.~\ref{fig:filter}.}
\label{fig:pd_combined}
\end{figure}

\end{example}

\newpage

\begin{figure}[h]
\centering
\includegraphics[width=1\linewidth]{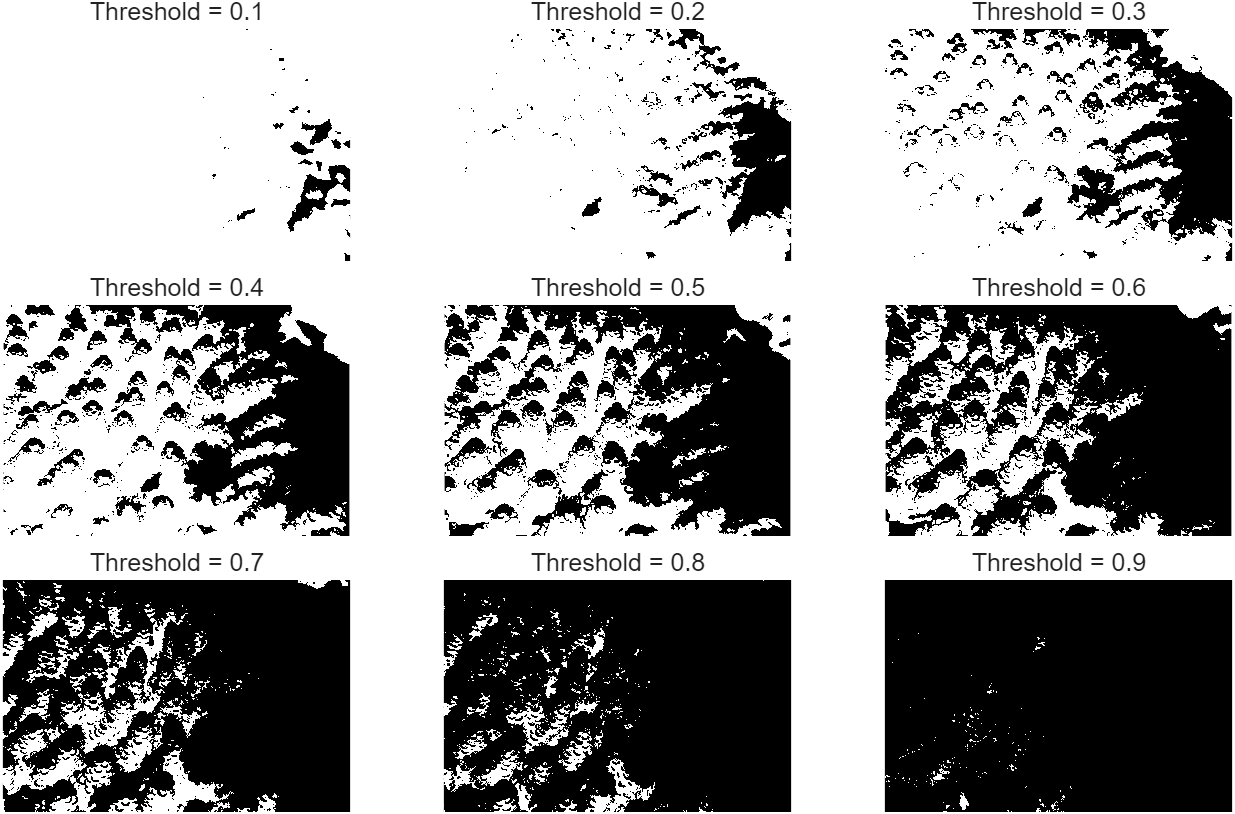}
\caption{Sliced image of \textit{Acropora nyinggulu} with thresholds moving from white = 0 to black = 1.}
\label{fig:slices}
\end{figure}

\begin{figure}[h]
\centering
\includegraphics[width=1\linewidth]{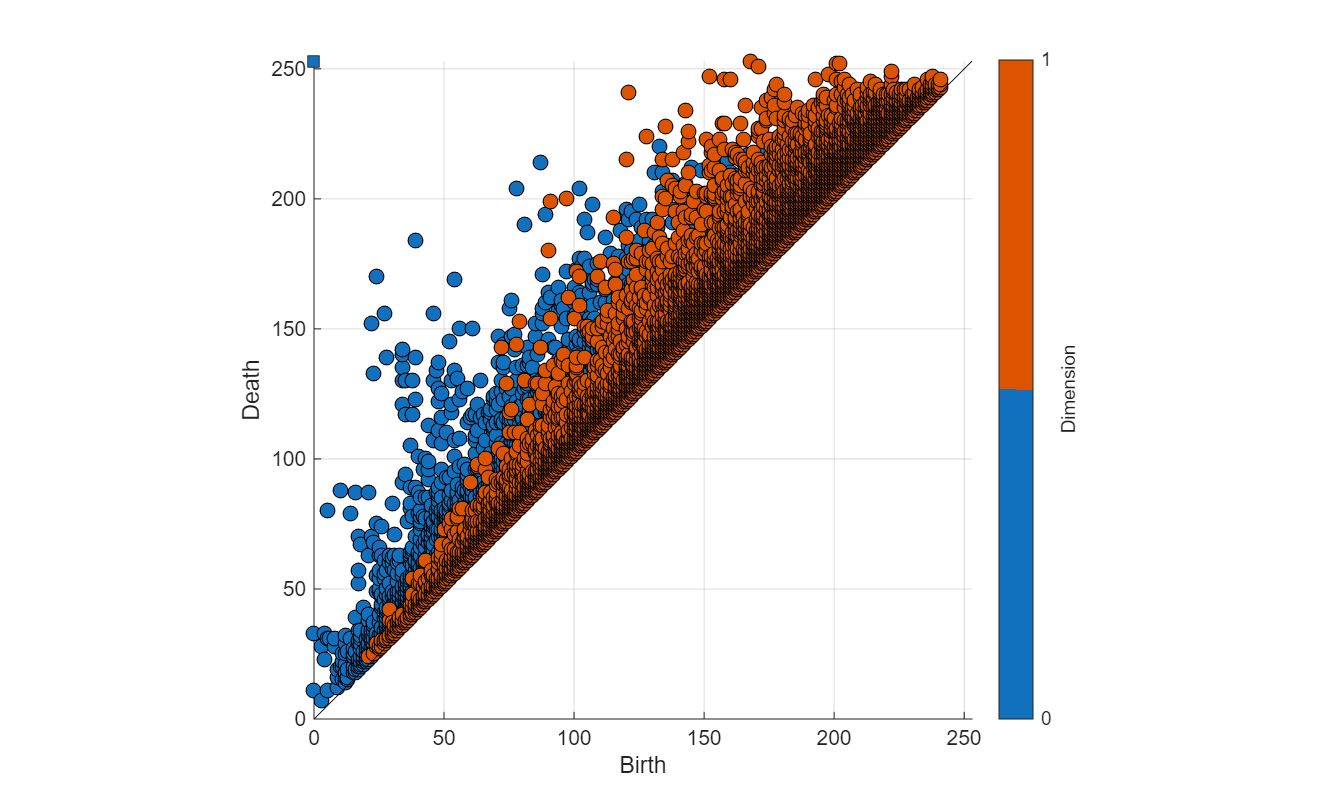}
\caption{Corresponding persistence diagram from the coral image used in Fig.~\ref{fig:slices}.}
\label{fig:diag}
\end{figure}

To simplify the process in our coral examples, Fig.~\ref{fig:slices} shows a coral image in black and white at different intensity thresholds. The first features to be classified as black are the tips of the coral branches. Then the outlines of the branches become clearer as the polyps, which stick out, turn black. The dense base of the coral, where the branches are attached, is filled in last. Of course, this simple schematic does not give us direct insight into quantifiable topological data, but it may make it easier to understand what we are looking for in the results. This mechanism allows for focus on features that tend to lie in the foreground. This process ensures that structurally important features of corals, such as the polyp arrangement around the branches, are emphasised. In Fig.~\ref{fig:diag}, the corresponding persistence diagram is displayed. Some general observations can be made about the shape and distribution of the data points, but it is not diagnostic.

\newpage

In our corals, for example, many loops appearing at similar birth values and dying at similar times may indicate a regular distribution of coral polyps. The persistence, or how long a feature remains distinct from its surroundings, often gives it more meaning. We expect features such as branch tips and edges to stand out prominently. That can give us information on branch density and when branches merge into the main coral colony. While we tend to focus on high-persistence features, lower-persistence features may provide information about surface texture. Unfortunately, low persistence can also be due to noise, making it more challenging to determine which feature classes indicate corals. Medium persistence might contain the most valuable information, likely focusing on the branch structure, so information across the coral for individual branches. The high persistence values likely correspond to the overall architecture of the coral colony, answering questions such as how compact the colony appears to be. The power of persistent homology, which combines small and large features, makes it an attractive tool for studying coral morphology. 

\subsection{Images}
We used raw images of tabular Acropora specimens published in the taxonomic revision by \cite{rassmussen2025tables}, most of which are housed in the collections of the Queensland Museum of Tropics (QMT). For each image we followed this protocol: firstly, 30 random patches were extracted from each image by fixing a centre point and specifying a given number of pixels in each direction, ensuring equal-sized patches across the image. The number of patches acts as a practical trade-off between sample size and computational cost, consistent with common heuristics for approximating normality in sample means. The random nature of patching allows for overlapping. Each patch contains 564,001 pixels (751 x 751 px) and is a *.png file. Afterwards, each patch was converted to grayscale, inverted, and normalised by dividing each pixel intensity value by the max value $I_{ij}/I_{max}$, resulting in a range of 0-1 for each patch. Then, the image was converted to and saved as a DIPHA-compatible format using the DIPHA-provided MATLAB function. The grayscale was inverted because DIPHA's filtration progresses from black to white; therefore, the inversion ensures that typical foreground features, such as branch tips, are considered first.
The images were pre-selected to minimise disturbance from low-quality and high-background proportions. Unfortunately, no scale was available to accurately compare the specimen size, but a similar proximity to the coral was considered.

\subsection{DIPHA}
The main software application used for this research is DIPHA, a distributed persistent homology algorithm \citep{dipha}. Gray-scale images are put into the algorithm, which produces persistence diagrams as an output. DIPHA was compiled from source using CMake within a Windows Subsystem for Linux (WSL) Ubuntu environment. DIPHA is particularly built to compute cubical complexes, rather than simplicial complexes. This makes it the preferred software for the available image data.
All other image processing, statistical analysis, and figure generation were performed in MATLAB R2025b.

\subsection{Rank Function}
Persistence diagrams are difficult to compare because they have different numbers of points. Summarising the information from the persistence diagram into a function makes it easier to use in statistical analyses. For this reason, we work with rank functions. The rank function is a function from $[0,1]^2 $ to $ \{0,1,2,…\}$. For points $(b,d)$ such that $b \leq d$, the rank function counts all the data points to the top and to the left of $(b,d)$. The persistence diagram only contains points where $birth < death$, so for all points below the diagonal $(b = d)$, we set the $rank = 0$.
So, at any point $(b,d)$, the rank function is the number of homology classes with $birth \leq a$ and $death \geq b$. It essentially counts the features that were already born by time $b$ and are still alive at time $d$.

\begin{figure}[htbp]
\centering
\includegraphics[width=\linewidth]{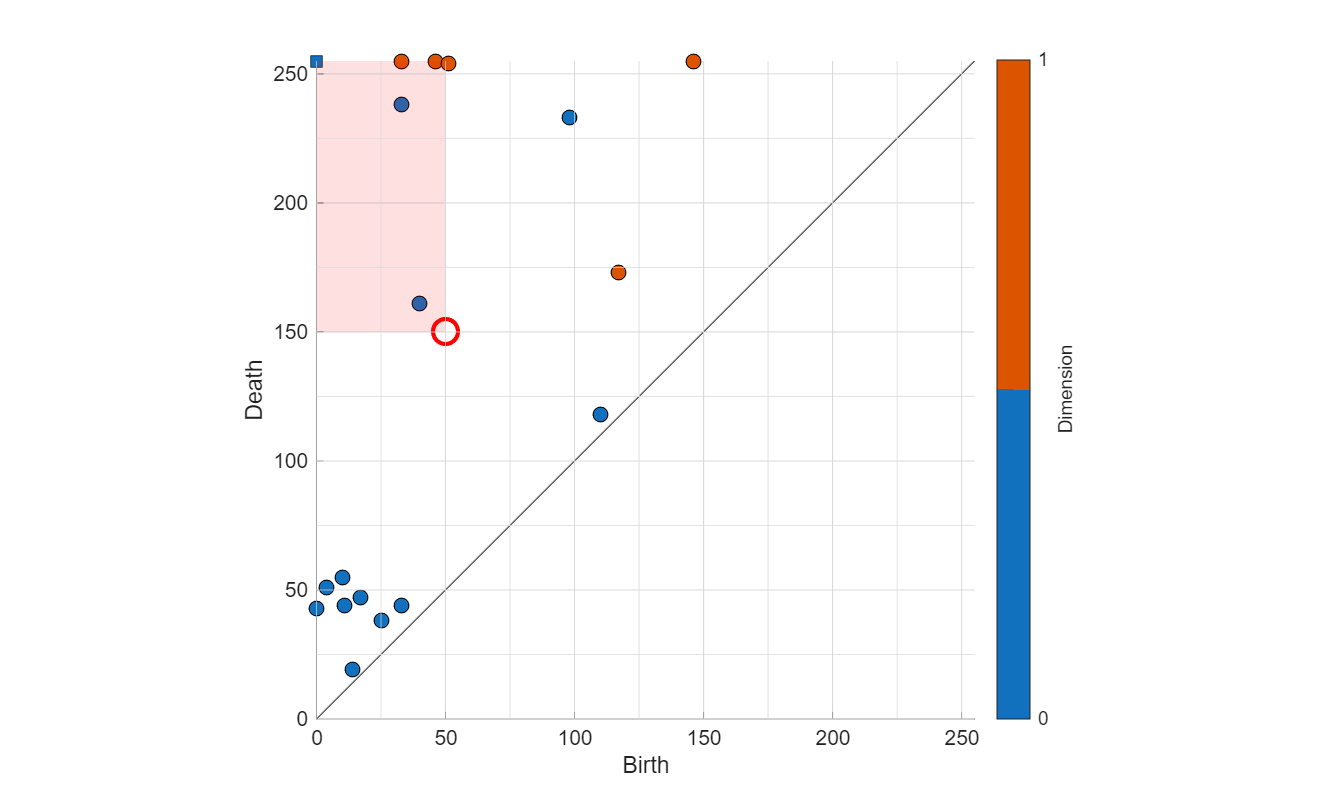}
\caption{Rank evaluation region for $R(50,150)$.}
\label{fig:shadedreg}
\end{figure}

As seen in Fig.~\ref{fig:pd_combined}, the shaded region visualises all the data points that are counted at the lattice point $(50,150)$ to evaluate the rank. This persistence diagram is the direct output from the filtration in Fig.~\ref{fig:filter}. It shows many connected components born early, also dying early (agglomeration of data points in the lower left corner). Whereas the loops born early all persist until the end of the filtration. The grid is sampled at 81 points: $a,b \in \{1,2,3,4,5,6,7,8,9\}$.
The resulting rank function containing both connected components and loops is
\[
R =
\begin{bmatrix}
1 & 4 & 4 & 4 & 4 & 5 & 5 & 5 & 5 & 5\\
1 & 5 & 5 & 6 & 6 & 7 & 7 & 7 & 7 & 0\\
1 & 5 & 5 & 6 & 6 & 7 & 7 & 7 & 0 & 0\\
1 & 5 & 5 & 6 & 6 & 7 & 7 & 0 & 0 & 0\\
1 & 6 & 6 & 7 & 8 & 9 & 0 & 0 & 0 & 0\\
1 & 6 & 6 & 7 & 8 & 0 & 0 & 0 & 0 & 0\\
1 & 6 & 6 & 7 & 0 & 0 & 0 & 0 & 0 & 0\\
1 & 6 & 6 & 0 & 0 & 0 & 0 & 0 & 0 & 0\\
3 & 8 & 0 & 0 & 0 & 0 & 0 & 0 & 0 & 0\\
6 & 0 & 0 & 0 & 0 & 0 & 0 & 0 & 0 & 0
\end{bmatrix}
\]

The map from persistence diagrams to rank functions is injective, meaning each rank function uniquely determines the persistence diagram from which it was computed, and the diagram can be reconstructed from it. When working with discretisations, we lose injectivity, but given a set of persistence diagrams there is always a discretisation of the rank functions which retains injectivity with respect to this set. To visually represent and compare the persistence diagrams, we overlay a grid on the persistence diagram with a given number of lattice points and only consider the rank at each lattice point. In technicality, the true rank function is defined for every real pair $(a,b)$ above the diagonal, so the rank function we compute based on the lattice points is a discretisation, not the full continuous function.

The rank function then returns a matrix that summarises the information in the persistence diagram. The grid ensures that we have the same number of lattice points for each persistence diagram, which means the rank function matrices will be the same size.   The rank functions can also be visualised as heatmaps, which show how many topological features are born at a given threshold (x-axis) and their associated lifetimes (y-axis). Areas of interest may include strong drops in lifetime over a short range of birth times, or the evenness of persistence across features born at the same time. 

\subsection{Principal Component Analysis}

In the next step, the mean of the rank functions across all patches per image was computed. This produces an average rank function summarising the persistence information for an entire image. The rank function matrices are then converted to vectors for Principal Component Analysis (PCA). This process follows the approach of \cite{robins2016pca}.  
The functions were centred using the median; they were also checked against the mean, showing stability under both centring methods. In this method, we also have the option to apply a persistence weight, which introduces a slight bias toward longer-persistence features, for example, highlighting them over noisy, short-lived features. The weight is usually applied to emphasise robust topological features over near-diagonal noisy attributes. 

In a last step, we perform PCA with the species labels assigned to our data points. They are plotted based on their PC scores. To sort the species in the function space, we apply ellipses based on the standard error of the mean (SEM). This results in elliptic areas, where we would expect the data points to belong to the same species.

\section{Results}
\subsection{Rank Function Observations}
To compare how persistent homology extracts different features from in-situ coral images, we first analyse the heatmaps. Here, two different heatmaps are examined. The heatmaps consist only of $H_1$ features (loops), so we do not need to consider how the ratio of loops to connected components may affect the total number of features across images. The scale and colourbar are the same across the different heatmaps, making them easier to compare visually. The heatmaps we are discussing were chosen based on the distribution pattern of the number of features. We are focusing on rather different patterns to explain where these differences may come from, and what image features are important in shaping the rank function. 
\begin{figure}[H]
\centering
\includegraphics[width=\linewidth]{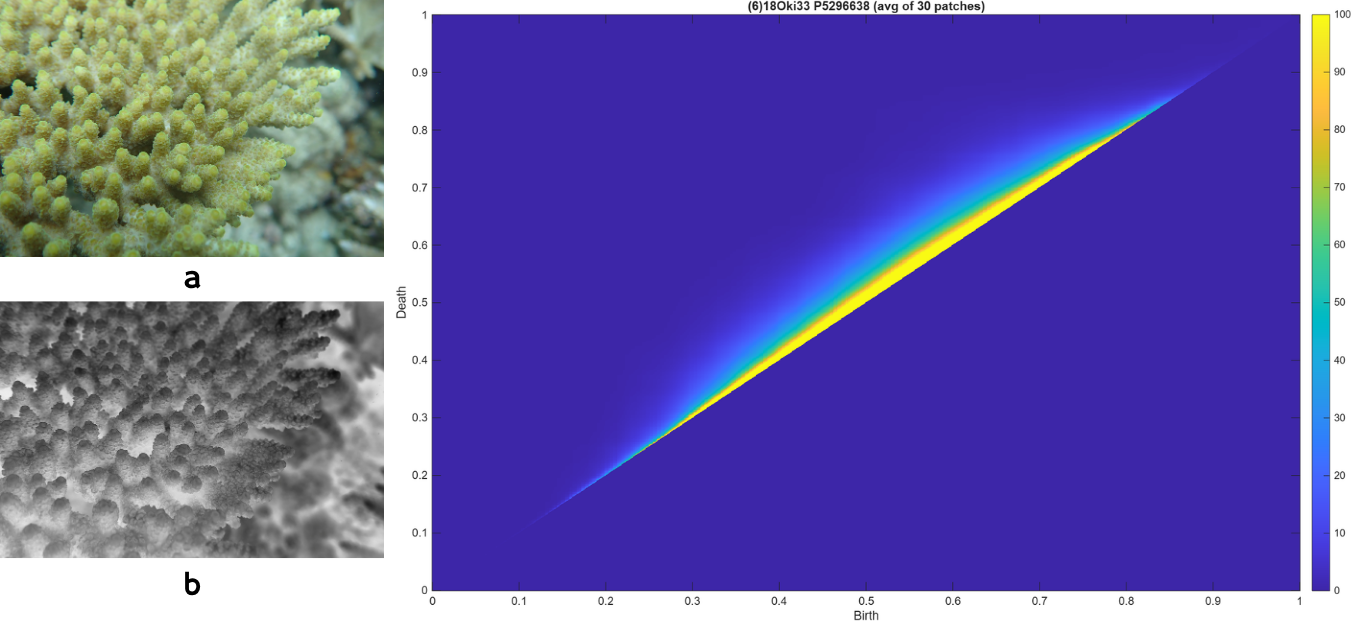}
\caption{Heatmap built from the average rank function with birth on the x-axis and lifetime on the y-axis. Colourbar corresponds with the number of features. Heatmap only considers loops.
\textbf{a)} Original image of \textit{A. bifurcata}. From \textit{QMT CoralBank}.
\textbf{b)} Corresponding inverse grayscale image.
}
\label{fig:img6}
\end{figure}

This heatmap shows a narrow, high-intensity band closely tracking the diagonal (birth = death) (Fig.~\ref{fig:img6}). Features begin to appear around birth $\approx$0.1, with no signal present in the lower-left region of the diagram. Features are present across most of the range between birth $\approx$0.3 and 0.8, with very few present after birth $\approx$0.85. The rank value is highest for features born around the midsection of this range, indicating that persistence is greatest there. Across the diagram, the band remains narrow throughout, with most features persisting for only a short interval — up to a persistence of $\sim$0.1, and typically around a quarter of that or less.

\begin{figure}[H]
\centering
\includegraphics[width=\linewidth]{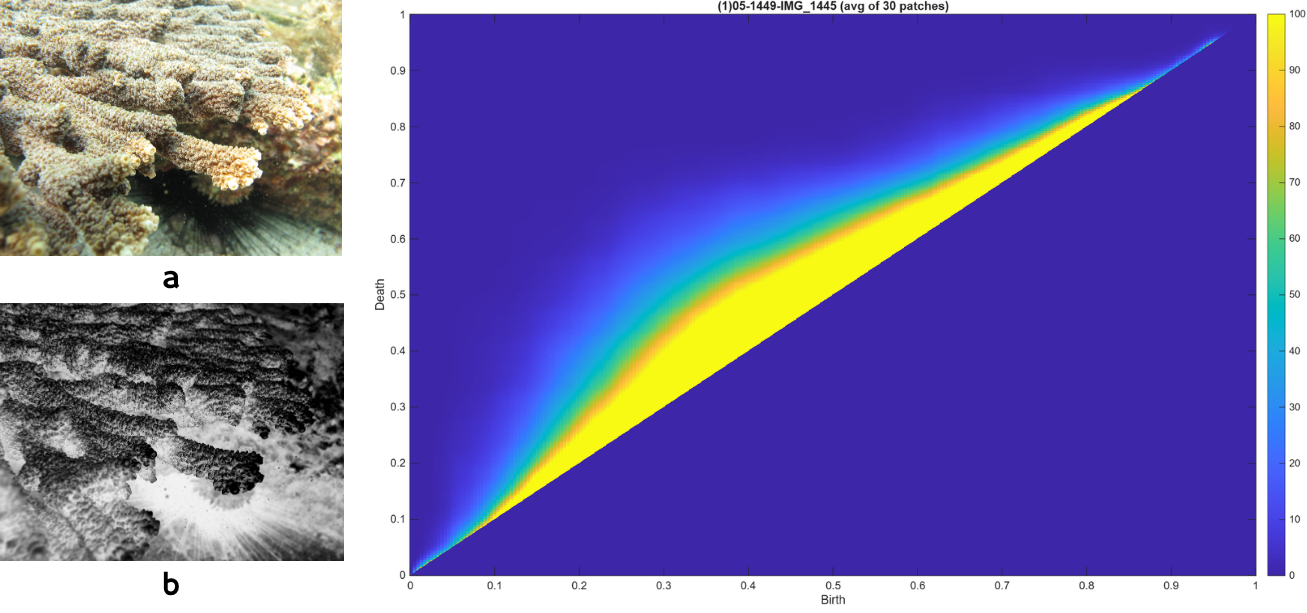}
\caption{Heatmap built from the average rank function with birth on the x-axis and lifetime on the y-axis. Colourbar corresponds with the number of features. Heatmap only considers loops.
\textbf{a)} Original image of \textit{A. sp.}. From \textit{QMT CoralBank}.
\textbf{b)} Corresponding inverse grayscale image.
}
\label{fig:img12}
\end{figure}
The high-intensity core of this heatmap follows a shallow, right-skewed bell-like shape, with both the bandwidth and rank value peaking around birth $\approx$0.35–0.5 (Fig.~\ref{fig:img12}). Features are present across nearly the entire range (0–1), with the majority occurring between birth $\approx$0.1 and 0.9. The highest rank values and longest persistence, reaching up to $\sim$0.3, are concentrated around this peak region, while the band narrows toward a single point near birth $\approx$0.95, death $\approx$0.97, indicating a sharp drop in both persistence and rank value at high birth values.

\subsection{Principal Component Analysis}
PC1 accounted for 66.13\% of the total variance, and PC2 accounted for 28.04\%. Specimens are mainly distributed along the PC1 axis, with PC1 scores ranging from -0.4 to 0.45, whereas PC2 scores range from -0.25 to 0.2. \textit{Acropora spicifera}, \textit{Acropora tersa}, and \textit{Acropora pectinata} are all positioned on a linear SEM, being restricted by only 2 specimens. \textit{Acropora downingi} shows a larger horizontal distribution with a PC1 range between -0.25 and 0.35, positioned in the right hemisphere of the plot. \textit{Acropora bifurcata} is the most scattered species, with the SEM only enclosing 1 out of the 5 specimens. The ellipse of \textit{Acropora nyinggulu} encloses 3 out of the 4 specimens; however, it is located inside the three ellipses of \textit{A. downingi}, \textit{A. hyacinthus}, and \textit{A. bifurcata}. 

\begin{figure}[h]
\centering
\includegraphics[width=1\linewidth]{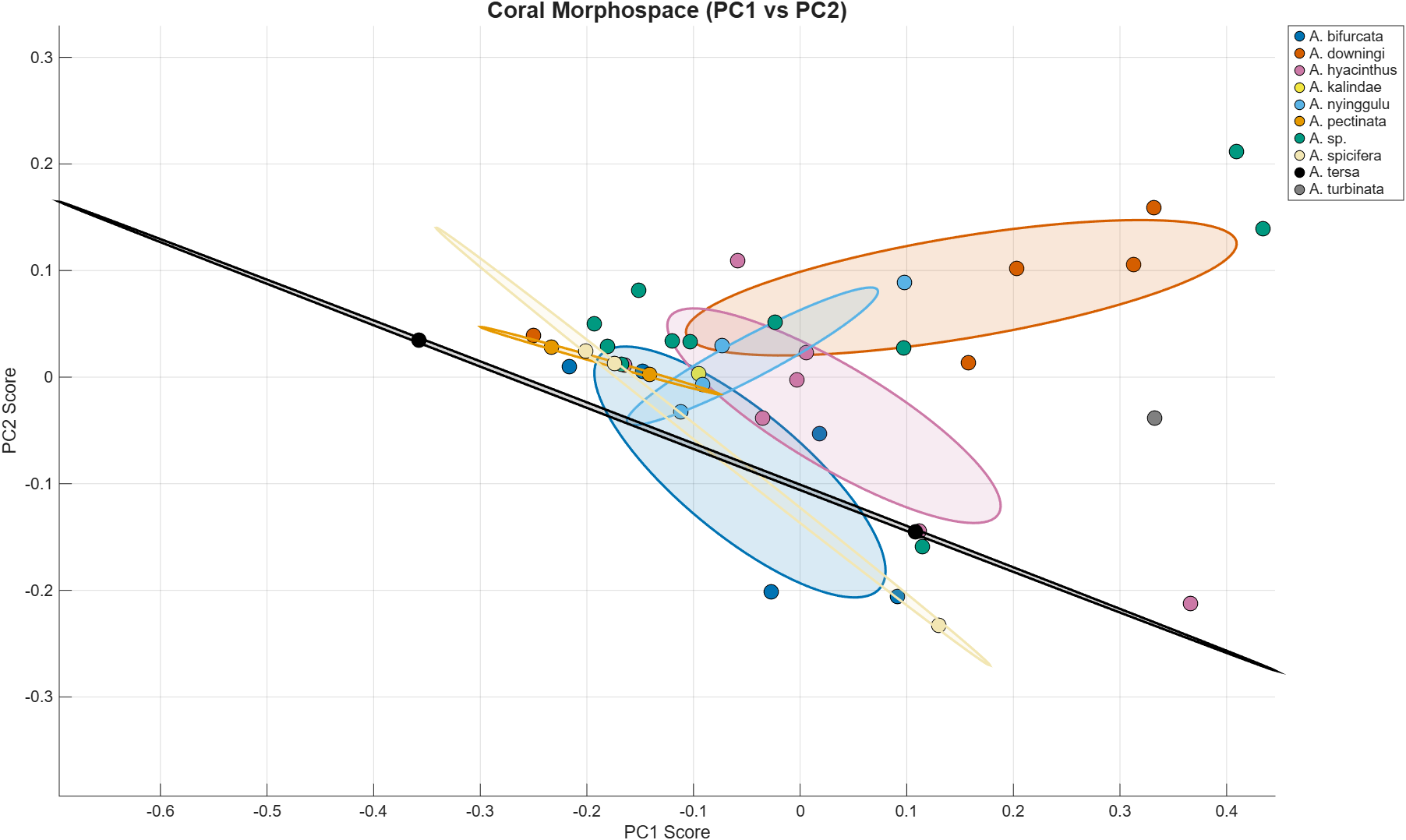}
\caption{PCA plot with supervised classification and ellipses displaying the 95\% standard error of the mean (SEM) for each species.}
\label{fig:morpho}
\end{figure}

\section{Discussion}
\subsection{Heatmaps}
The heatmap (Fig.~\ref{fig:img6}) contains a narrow band of rank values sitting close to the diagonal. The corresponding photograph for this specimen (Fig.~\ref{fig:img6}) shows a densely packed arrangement of narrow and short, closely spaced branches with relatively uniform gaps between corallites. This structural regularity is consistent with the narrow rank-function band observed in Fig.~\ref{fig:img6}. Because the spacing between branches is fairly consistent across the imaged surface, topological features (gaps between branch structures) tend to form and close within a similarly narrow range of filtration values, producing a tightly concentrated band rather than a broad spread. The short persistence of these features (rarely exceeding~0.1) further suggests that the gaps captured by the filtration are relatively small and fill in quickly, consistent with the fine, closely-packed branching visible in the photograph, rather than reflecting larger-scale structural voids in the colony.

In the second heatmap (Fig.~\ref{fig:img12}) we see a much thicker band of rank values with higher intensity. The corresponding photograph (Fig.~\ref{fig:img12}) shows a flatter branching structure, with irregular spacing between elongated branches of varying lengths. This structural variability is consistent with the considerably broader rank-function band observed for this specimen (Fig.~\ref{fig:img12}). Topological features form and close across a much wider range of filtration values than in the more uniform specimen shown previously. The higher persistence values observed here (reaching up to~0.3, compared to~0.1 for the previous specimen) likely reflect the strong contrast between the top of the branches and the very light background. Small corallites across the coral surface stand out significantly, consistent with the longer persistence for many features.

In general, the interpretation of these heatmaps may not directly indicate morphological differences that are valuable for species differentiation. However, they act as a first visible quantification of topological features within our image data. This step supports the biological interpretation of persistent homology by providing a visual aid. Many of the topological features identified by DIPHA are not visible to the naked eye, which is why an analysis like this offers a completely new perspective on coral morphology. Yet, these heatmaps are crucial for explaining how this method works and for identifying which images may be more or less suitable for this analysis. In particular, they can determine whether non-coral structures, such as background substrate, influence the results. This can provide information for future data collection or how images should be pre-processed to exclude non-coral structures. Another important advantage of analysing heatmaps is the identification of significant areas within the persistence diagram. In return, this information can help design weight functions for PCA that may emphasise the features within these areas of significance.

\subsection{PCA and Comparison of Original Images}

\begin{figure}[H]
\centering
\includegraphics[width=\linewidth]{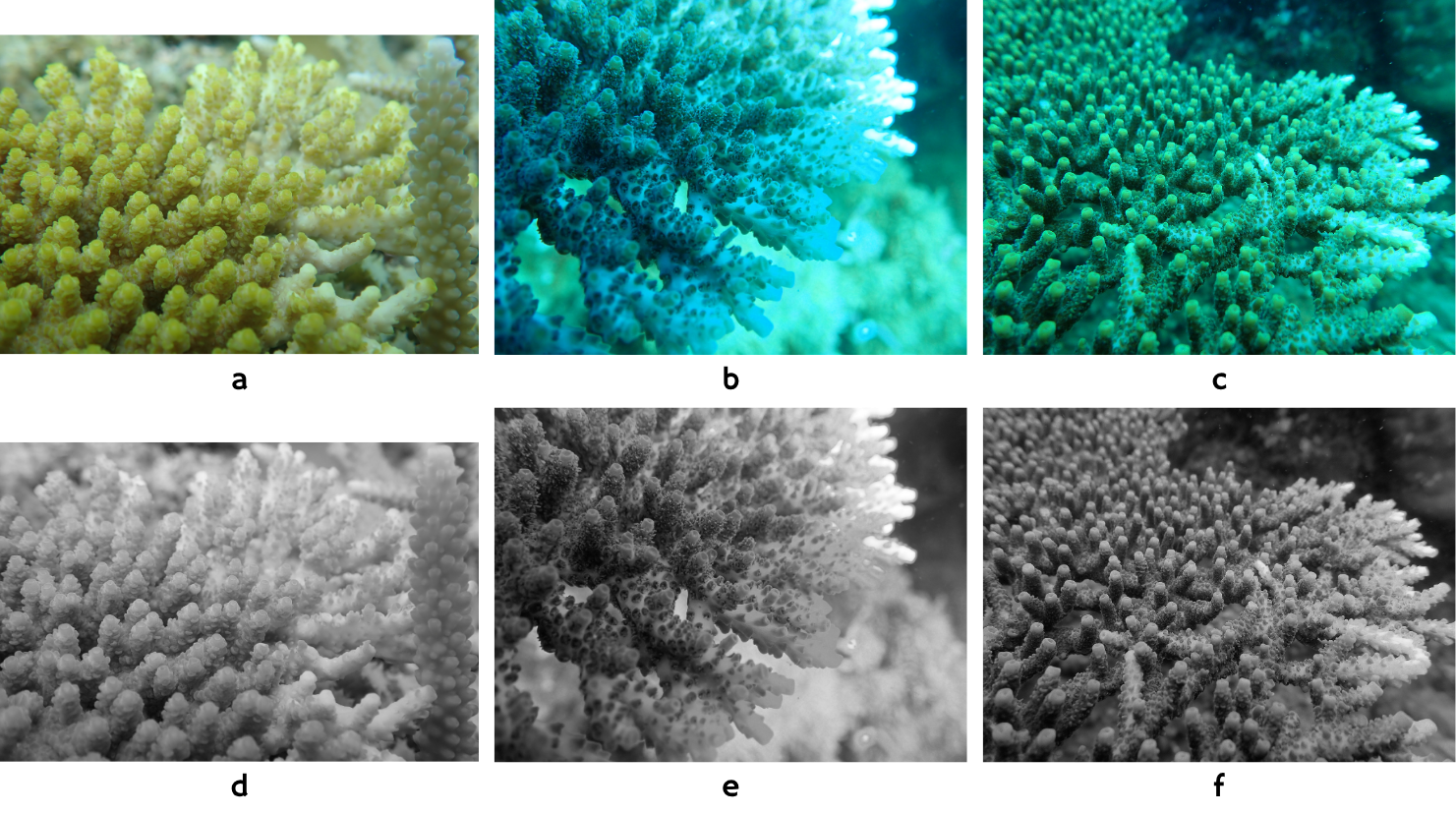}
\caption{Three different coral images in colour and grayscale of the same species \textit{A. bifurcata}. From \textit{QMT CoralBank}.}
\label{fig:13etc}
\end{figure}

The 2D PCA plot explains 94.17\% of the variance in the rank function of the coral specimens. The general horizontal spread suggests that the first principal component largely drives the spatial distribution. While none of the ellipses are discretely separated, they provide a useful visual indication of within-species variability. To further illustrate how differences in the rank function relate to species-level morphology, representative heatmaps and corresponding coral photographs are provided in Supplementary Material S1. \textit{A. bifurcata} may have a larger discrepancy between the PC scores; the specimen images of this species also show high differences. In Fig.~\ref{fig:13etc}, all three images are different hues. Image \textbf{a} is white-balanced, eliminating the typical blue and green colours from underwater photography. Image \textbf{b} is mostly blue, whereas image \textbf{c} is impacted by green hues. 

\newpage

\begin{figure}[H]
\centering
\includegraphics[width=\linewidth]{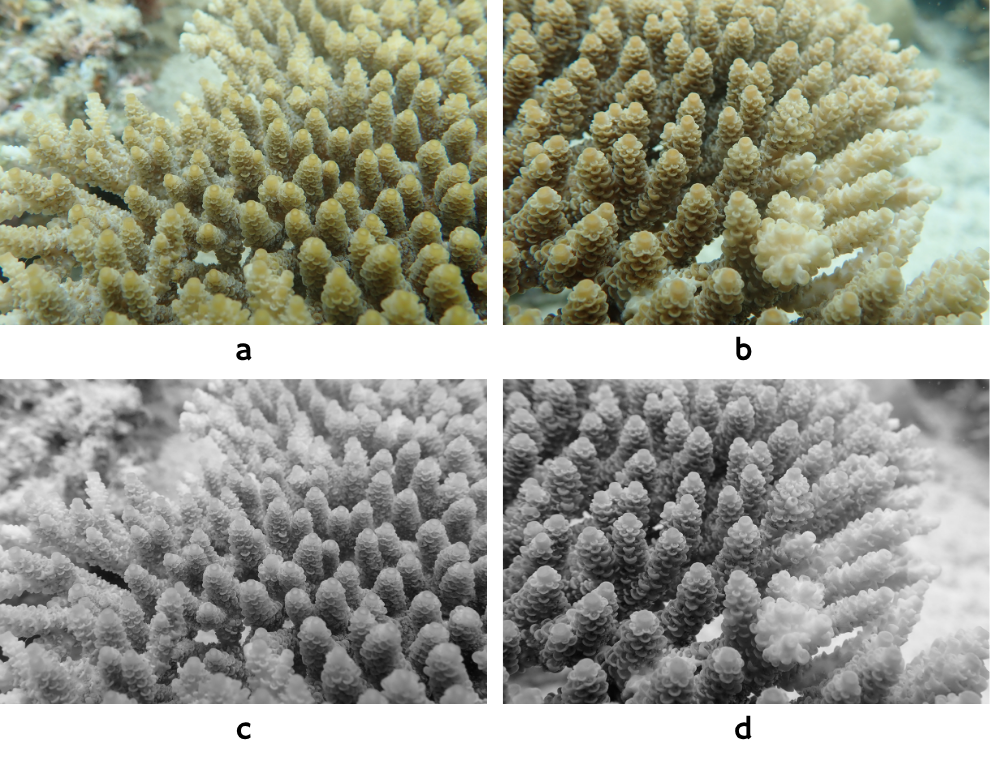}
\caption{Two different coral images in colour and grayscale of the same species \textit{A. nyinggulu}. From \textit{QMT CoralBank}.}
\label{fig:4and5}
\end{figure}
In image \textbf{c/f}, the tips of the branches are the most distinct. Image \textbf{d} seems to have a more uniform light gray colouring around the branches, whereas image \textbf{e} is much darker (except for the background), and emphasises the living polyps protruding from the coral. Image \textbf{e} also has lower resolution around the coral's edge, which may inhibit topological features from being picked up by the algorithm.   
In the PCA plot, there are two coral specimens belonging to the \textit{A. nyinggulu} that have almost identical PC scores. Images \textbf{a/c} and \textbf{b/d} almost appear to be the same image but mirrored (Fig.~\ref{fig:4and5}). Both images show a similar background-to-foreground ratio and feature the coral's edge, thereby capturing branches at different angles. While the tips of the branches are not as emphasised as in other examples, both images feature very distinct corallites adorning the branches.

\begin{figure}[H]
\centering
\includegraphics[width=\linewidth]{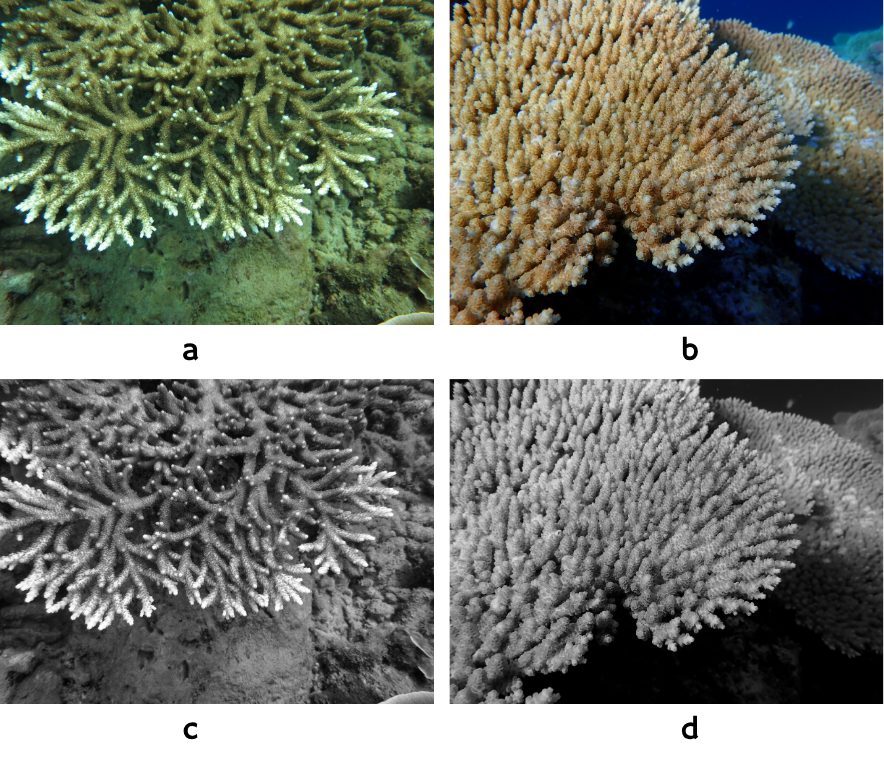}
\caption{Coral images in colour and grayscale of the two species \textit{A. sp.} (\textbf{a} and \textbf{c}) and \textit{A. hyacinthus} (\textbf{b} and \textbf{d}). From \textit{QMT CoralBank}.}
\label{fig:9and37}
\end{figure}

Another interesting example, is image \textbf{a/c} and \textbf{b/d} (Fig.~\ref{fig:9and37}). Image \textbf{a/c} is not identified to the species level but is closest to image \textbf{b/d} (\textit{A. hyacinthus}) in the PCA plot. Image \textbf{a} contains less compact branching, focused on outward expansion and less vertical growth. The tips of the branches are marked by significant white dots, and the living polyps are distinctly darker than the branches beneath. The image features a wide coral edge and a chunk of background. In contrast, image \textbf{b} features much shorter branches, all clearly angling upwards. There is a similar area covered by the coral's edge, and some background is visible. The tips of the branches are not distinctly coloured at all. The only similarity to the naked eye appears to be the prominent dark-coloured polyps covering the branches. Furthermore, looking at the coloured images, image \textbf{a} has a greenish hue, whereas image \textbf{b} is colour-corrected. Considering both images feature pronounced polyps, it seems the colour difference did not affect the grayscaling noticeably. This example supports the idea that polyp arrangement could be much more important in shape-based categorisation of corals than previously believed.

\subsection{Limitations}
This analysis was intended as a low-cost, efficient first trial of GTDA's potential for analysing coral images. By using a user-friendly open-source GTDA software and existing in-situ coral images, we were able to keep resource use minimal whilst gaining vital insights into how this research could be refined. As such, there are general limitations to this study, which motivate reassessing the study design with our findings in mind.

First, the lack of metadata should be noted. There is no way to compare the sizes of the coral specimens and normalise for proximity to the subject without a scale. Some images contain way fewer branches than others, and it is not always clear whether that is a morphological distinction or a photographic choice. Another varying factor is the lighting across the images. While some samples are colour-corrected, others are affected by the light absorption, leaving blue and green colour casts. It is also unclear how much the camera angle affects the algorithm's selection of certain topological features. Since we are analysing 2D projections of 3D objects, the silhouette changes with different angles. A branching coral photographed from above may appear highly connected, whereas the same colony photographed side-on may produce more loops ($H_1$) due to visible gaps between branches.  Both camera handling and lighting can also cause highlights and shadows that disrupt the grayscale. The images additionally had varying resolutions, making most unsuitable for further analysis. At lower resolutions, fine-scale morphological features such as individual polyps are not visible, resulting in entire branches appearing as uniform grey regions. This loss of detail reduces tone variability and limits the detection of topological features. Even in the image examples, some lose focus at the edges, which in turn affects the topology. While efforts were made to minimise the disturbance of these factors by pre-selecting images that visually appeared to be of better quality, it is suggested that further investigation be undertaken. It is important to note that the images used were not specifically taken for this study, and therefore, the photographers did not consider these issues. It is an efficient and low-cost way to research new methods, yielding parameters that can optimise future data collection. 

It is also not yet clear how strongly the proportion of background structures, such as the substrate, may influence the topological features. While most of them were chosen with that question in mind, it is not fully possible to eliminate all non-coralline structures in each image. Given that the PH process biases foreground features and can be further strengthened by PCA weightings, the background is unlikely to interfere strongly with the results; however, no quantitative certainty can be assigned. 

\subsection{Future Directions}
Considering this method is a very first attempt at incorporating GTDA into the coral taxonomy space, it is to be expected that there are many possible future trajectories. In the next step, we would hope to emphasise the differentiation between the homology feature classes $H_1$ and $H_0$. Rather than including both in the persistence diagrams and treating the dataset as a single entity in the analysis, we would advise repeating the process for each class separately. Focusing only on the loops may distinguish morphologically intricate areas of the coral. The proportions of $H_1$ to $H_0$ dimensions may also vary across species, improving the ability to identify which features may indicate significant morphological differences.

Another step is to ideally increase the sample size. After selecting images based on resolution, coral-to-background proportions, and their proximity to the coral, only 41 images remained for this study. Of those 41, 11 specimens were not identified to the species level. Obviously, the low number of images per species minimises the statistical significance of the results. If a larger number of specimens supports the current SEM ellipses, the proposed PCA may already be an additional tool that can help narrow down the possible taxonomy of the unidentified \textit{Acropora}. With more images of varying quality, it might also be possible to identify specific ways to account for image variation. So, finding ways to standardise for different resolutions, for example. While it may seem paradoxical to use lower-grade images, it can help us understand how strongly the method is affected by the image quality and whether there are simple techniques to account for these photographic differences.

The data folder used in this study also includes skeletal images. It would be interesting to see whether similar patterns to those of our in-situ corals appear in the PCA plot. Perhaps skeletal images are better suited to this method, as they are not disrupted by lighting changes that alter image colour or introduce shadows. Of course, in the long term, it is not sustainable to collect coral skeletal fragments each time a species ID is requested. However, with the images already available, it can certainly help to understand what image features are important for a successful GTDA analysis.

When analysing this method statistically, it may be of interest to use the accumulated persistence function rather than the average rank function. The average rank function preserves information from the persistence diagram by counting the number of persistent features at each threshold, then averaging across multiple image patches. The accumulated persistence function sums up the persistence of all features up to a given threshold \citep{biscio2019accumulated}, thus, naturally biasing long-lived features and stressing importance rather than abundance. Therefore, testing the accumulated persistence function in coral morphology may further help identify which topological features are important in separating species.

Another interesting point is to replace the PCA with a Convolutional Neural Network (CNN); however, this would require substantially larger datasets for training and validation. PCA was chosen as it is a more intuitive and interpretable tool. Furthermore, scientists working in marine biology and ecology are likely more familiar with PCA. As this method aims to introduce GTDA into the marine science space, it makes sense to use a tool that produces a somewhat visually interpretable output for simplification. However, CNNs do have advantages that cannot be ignored. CNNs automatically learn hierarchical features from data, which makes pattern recognition more accessible \citep{prahalathan2024improving, vijaiyan2023enhanced}. This process could be very useful for studying coral morphology and for more easily identifying species-specific differences. There are, of course, various other tools to be explored in this context, and new methods are continually being developed \citep{wasserman2016tda, carlsson2009topology}. \cite{kanari2017topo}, for example, developed the Topological Morphology Descriptor, applicable to "tree-shaped" branching structures, which may be useful for \textit{Acropora} with prominent branches.

In the long term, we hope to expand this method to other coral genera. While \textit{Acropora} are the main focus and of interest in reef conservation, it would be appealing to see how the topological features will vary in other growth forms. Especially, corals that do not exhibit any branching and have less obvious morphological differences would be of interest. Expanding the study to these growth forms would also shed light on which morphological markers are species-specific, as it goes beyond macromorphological appearance.

\subsection{Robustness}
During each step of this process, multiple data modifications were tested to ensure this method is robust. Patch sizes were chosen at three levels: small (375x375 pixels), medium (751x751 pixels), and large (1501x1501 pixels). The number of data points per patch across the three sizes was extracted and statistically analysed to assess normality, yielding similar patterns and trends. 

PCA centring was tested using both the median and the mean, and the swap did not affect the analysis outcome. Furthermore, PCA was conducted on individual patches to see how the deconstructed image would persist under different weightings. The weightings tested were log, power, exponential, negative exponential, inverse, and none. 

The clustering choice for unsupervised PCA was crudely tested. The Euclidean distance and Ward linkage were used in this method; however, different choices were applied and visually compared to ensure that clustering would not deviate significantly across different distance and linkage choices. The distances tested were: city block, Mahalanobis, cosine, and correlation. The linkages tested were: average, complete, single, and centroid. The default produced the best visual result; there were no significant differences between the clustering choices in the PCA plot. These clustering choices would likely be more important for producing a dendrogram than for a PCA plot. In coral taxonomy in particular, the dendrogram would be directly comparable to existing phylogenetic trees. In this study, the dendrogram was not very useful as the species did not cluster together perfectly.

\subsection{Conclusion}
This study represents an early step towards integrating mathematical tools with biological interpretation to inform coral taxonomy. While the PCA captured the majority of variance in the rank-function representation of coral morphology (94.17\% across the first two principal components), most species showed considerable overlap in their confidence ellipses, reflecting the continuous and often subtle morphological variation characteristic of the genus \textit{Acropora}. Despite this overlap, some contrasting distribution patterns of species suggest that persistent homology may be capable of detecting genuine morphological distinctions. This result indicates that further refinement of the method may improve its capacity to resolve species boundaries that have proven difficult to delimit using traditional morphological approaches. More broadly, this study highlights the potential value of mathematical and topological tools in biological research, and suggests that such approaches may complement existing methods in resolving longstanding taxonomic uncertainty within \textit{Acropora}.

\section{Acknowledgements}
We would like to thank James Cook University for making this project possible and for their student resources. This research was partially funded by the Austrian Science Fund (FWF) 10.55776/ESP9584724. For open access purposes, the author has applied a CC BY public copyright license to any author-accepted manuscript version arising from this submission. We furthermore wish to thank the Mathematical Research Institute MATRIX in Australia, where part of this research was conducted. We also acknowledge Andrew Baird and Sage Rassmussen for taking the images of the specimens used in this analysis and published in \cite{rassmussen2025tables}.

\bibliography{sample}

\end{document}